\documentclass[galaxies,article,accept,moreauthors,pdftex,10pt,a4paper]{mdpi}
\firstpage{1} 
\articlenumber{x}
\doinum{10.3390/------}
\pubvolume{xx}
\pubyear{2018}
\copyrightyear{2018}
\history{Received: date; Accepted: date; Published: date}

\usepackage{subfigure}
\usepackage{graphicx}

\usepackage{booktabs} 
\usepackage{multirow}
\usepackage{soul} 
\usepackage{microtype}
\usepackage{upgreek}

\Title{The PoGO+ Balloon-Borne Hard X-ray Polarimetry~Mission}

\Author{Mette Friis $^{1,2}$, M\'ozsi Kiss $^{1,2}$\orcidA{}*, Victor Mikhalev $^{1,2}$\orcidB{}, Mark Pearce $^{1,2}$\orcidC{} and Hiromitsu Takahashi $^{3}$\orcidD{},
on behalf of the PoGO+ Collaboration}
\address{%
$^{1}$ \quad KTH Royal Institute of Technology, Department of Physics, 106 91 Stockholm, Sweden; mettef@kth.se (M.F.); mikhalev@kth.se (V.M.); pearce@kth.se (M.P.)\\
$^{2}$ \quad The Oskar Klein Centre for Cosmoparticle Physics, AlbaNova University Center, 106 91 Stockholm, Sweden; hirotaka@astro.hiroshima-u.ac.jp\\
$^{3}$ \quad Hiroshima University, Department of Physical Science, Hiroshima 739-8526, Japan}

\corres{Correspondence: mozsi@kth.se; Tel.: +46-73-765-24-50}

\abstract{The PoGO mission, including the PoGOLite Pathfinder and PoGO+, aims to provide polarimetric measurements of the Crab system and \mbox{Cygnus X-1} in the hard X-ray band. Measurements are conducted from a stabilized balloon-borne platform, launched on a 1 million cubic meter balloon from the Esrange Space Center in Sweden to an altitude of approximately \mbox{40 km}. Several flights have been conducted, resulting in two independent measurements of the Crab polarization and one of \mbox{Cygnus X-1}. Here, a review of the PoGO mission is presented, including a description of the payload and the flight campaigns, and a discussion of some of the scientific results obtained to date.}

%keyword 1; keyword 2; keyword 3 (list three to ten pertinent keywords specific to the article, yet reasonably common within the subject discipline.)}

\begin{document}
%%%%%%%%%%%%%%%%%%%%%%%%%%%%%%%%%%%%%%%%%%
%% Only for the journal Gels: Please place the Experimental Section after the Conclusions

%%%%%%%%%%%%%%%%%%%%%%%%%%%%%%%%%%%%%%%%%%
%\setcounter{section}{-1} %% Remove this when starting to work on the template.
%\section{How to Use this Template}
%The template details the sections that can be used in a manuscript. Note that the order and names of article sections may differ from the requirements of the journal (e.g. the positioning of the Materials and Methods section). Please check the instructions for authors page of the journal to verify the correct order and names. For any questions, please contact the editorial office of the journal or support@mdpi.com. For LaTeX related questions please contact Janine Daum at latex-support@mdpi.com.
%The order of the section titles is: Introduction, Materials and Methods, Results, Discussion, Conclusions for these journals: aerospace,algorithms,antibodies,antioxidants,atmosphere,axioms,biomedicines,carbon,crystals,designs,diagnostics,environments,fermentation,fluids,forests,fractalfract,informatics,information,inventions,jfmk,jrfm,lubricants,neonatalscreening,neuroglia,particles,pharmaceutics,polymers,processes,technologies,viruses,vision

%%%%%%%%%%%%%%%%%%%%%%%%%%%%%%%%%%%%%%%%%%

\section{Introduction}
PoGO+ is a balloon-borne hard X-ray Compton polarimetry telescope operating in the energy range \mbox{$\sim$20--180 keV}. ``PoGO'' stands for ``Polarized Gamma-ray Observer,'' and the plus signifies an upgrade~\cite{PoGOLite to PoGO+}, specifically from the PoGOLite ``Pathfinder''~\cite{Payload paper}, which was launched in 2013~\cite{PoGOLite flight paper}. The~instrument is optimized for point-sources and has a field of view collimated to about \mbox{2$^{\circ}$}. Observational targets are the X-ray-bright sources on the northern hemisphere: the Crab and \mbox{Cygnus X-1}. 

By providing two new observational parameters, namely the polarization fraction and polarization angle, polarimetry is expected to provide new insight into the emission mechanisms, magnetic fields, and geometries of sources observed.~The relevance of polarimetric studies has been extensively discussed~\cite{These proceedings}, and results have been reported in particular for the Crab system. Many of the results suffer, however, from systematic effects, leaving open questions about the emission mechanisms at work in the observed astrophysical sources. The goal of the PoGO mission is to provide statistically constrained polarimetric observations (a minimum detectable polarization of \mbox{$\sim$10\%} before background subtraction), in a previously unexplored energy range, with well-understood systematic effects, using~an instrument that has been extensively calibrated with both polarized and unpolarized beams~\cite{PoGO+ calibration paper}.

Here, developments of the PoGO mission are collected, leading up to the 2016 flight and a summary of observational results for the Crab and \mbox{Cygnus X-1}. Section~\ref{Payload} gives a description of the PoGO+ payload, including the instrument, attitude control system, and gondola structure. A brief review of the flight campaigns to date, and the lessons learned from these, is provided in Section~\ref{PoGO flight campaigns}. In~Section~\ref{Science}, scientific results are presented. Finally, Section~\ref{Outlook} provides an outlook on some of the studies that are currently on-going using data collected with PoGO+.

%The introduction should briefly place the study in a broad context and highlight why it is important. It should define the purpose of the work and its significance. The current state of the research field should be reviewed carefully and key publications cited. Please highlight controversial and diverging hypotheses when necessary. Finally, briefly mention the main aim of the work and highlight the principal conclusions. As far as possible, please keep the introduction comprehensible to scientists outside your particular field of research. %Citing a journal paper \cite{ref-journal}. And now citing a book reference \cite{ref-book}. Please use the command \citep{ref-journal} for the following MDPI journals, which use author-date citation: Administrative Sciences, Arts, Econometrics, Economies, Genealogy, Humanities, IJFS, JRFM, Laws, Religions, Risks, Social Sciences.

%%%%%%%%%%%%%%%%%%%%%%%%%%%%%%%%%%%%%%%%%%

\section{The PoGO+ Payload}
\label{Payload}
\begin{figure}[H]
\centering
\includegraphics[width=.7\textwidth]{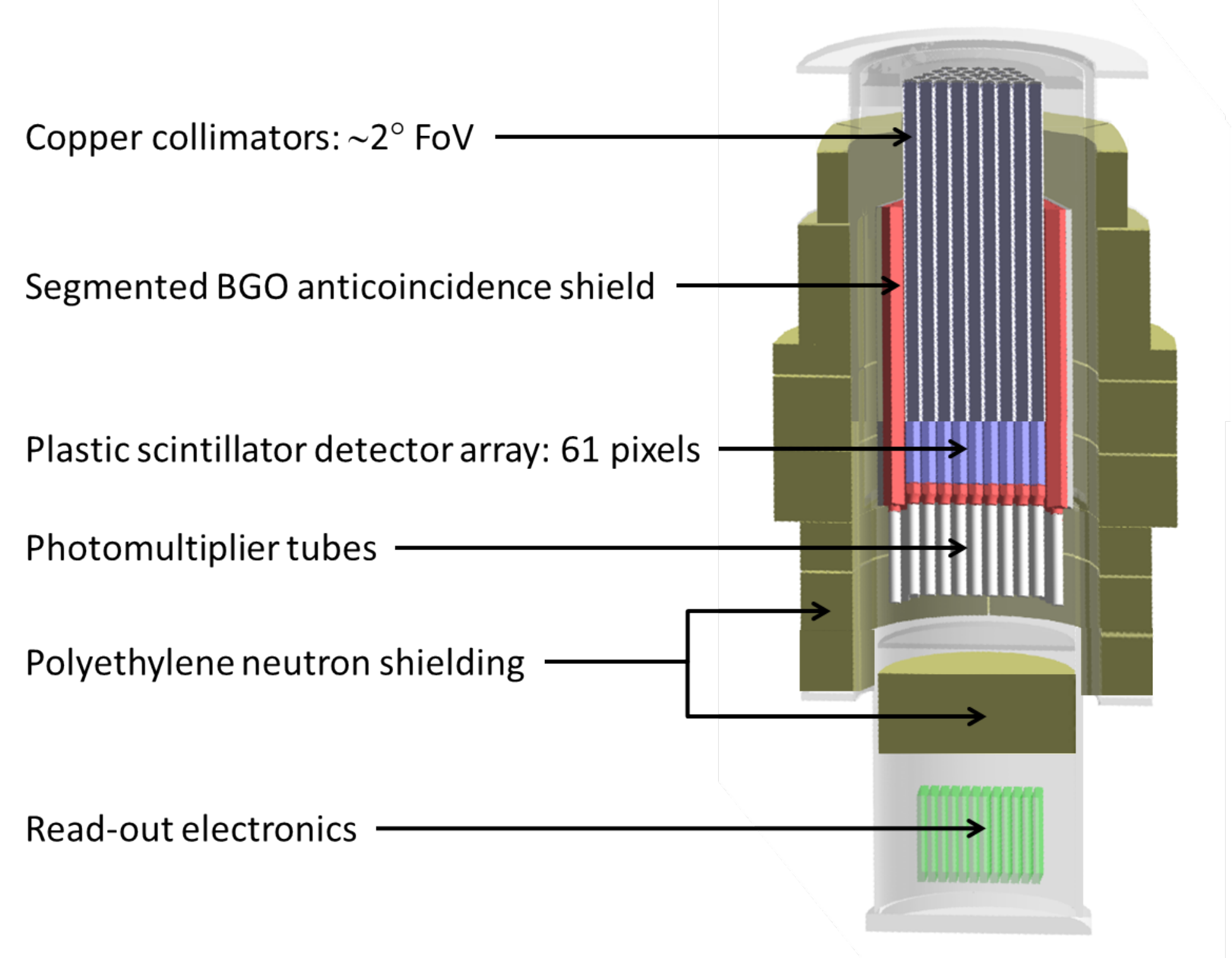}
\caption{\label{Detector sketch}Sketch of the PoGO+ polarimeter. The height is \mbox{$\sim$1.5 m}, and the total weight is \mbox{$\sim$600 kg}.}
\end{figure}

PoGO+ uses a 61-pixel array of plastic scintillators for reconstructing azimuthal scattering angles from two-unit interactions (polarization candidate events) in temporal coincidence.~As per the Klein--Nishina formula, such scattering angles will be modulated by the polarization of the incident photon beam, following a sinusoidal distribution with a \mbox{180$^{\circ}$} period~\cite{Lei 1997}.~The scintillators, type \mbox{EJ-204} from Eljen Technology, are \mbox{12 cm} tall and about \mbox{3 cm} across, with a hexagonal cross section, allowing them to be tightly packed in a honeycomb structure. Copper tubes, approximately \mbox{70 cm} in length, are used to collimate the field of view.~These are made from \mbox{0.5 mm} thin copper sheets, folded into hexagonal cross-section tubes, which are wrapped in \mbox{200 $\upmu$m} of tin foils followed by the same thickness of lead foils. This ``graded shield'' construction of the collimators efficiently suppresses the out-of-aperture background of the measurements. The plastic scintillators are surrounded on the bottom and on the sides by a bismuth germanium oxide (BGO) anticoincidence shield, segmented into 61 pieces on the bottom of the instrument (one corresponding to each plastic scintillator) and 30 pieces on the side.~The anticoincidence shield is \mbox{60 cm} tall (compared to the \mbox{12 cm} height of the plastic scintillators) and the thickness exceeds \mbox{3 cm} on each side. Plastic scintillators and side anticoincidence shield elements are read out by the same type of photo-multiplier tube, a modified version of Hamamatsu \mbox{R7899}. In order to cancel out variations in response between individual detector elements, the entire detector array rotates during observations at a rate of \mbox{1$^{\circ}$/s}. The~dominant background is from atmospheric neutron interactions~\cite{Kole neutrons}, wherefore the entire detector array is surrounded by a passive polyethylene shield, with a thickness between \mbox{5} and \mbox{15 cm}.~Figure~\ref{Detector sketch} illustrates the detector design. A more comprehensive description of the payload is presented in~\cite{Payload paper}.

\begin{figure}[H]
\centering
\includegraphics[width=.6\textwidth]{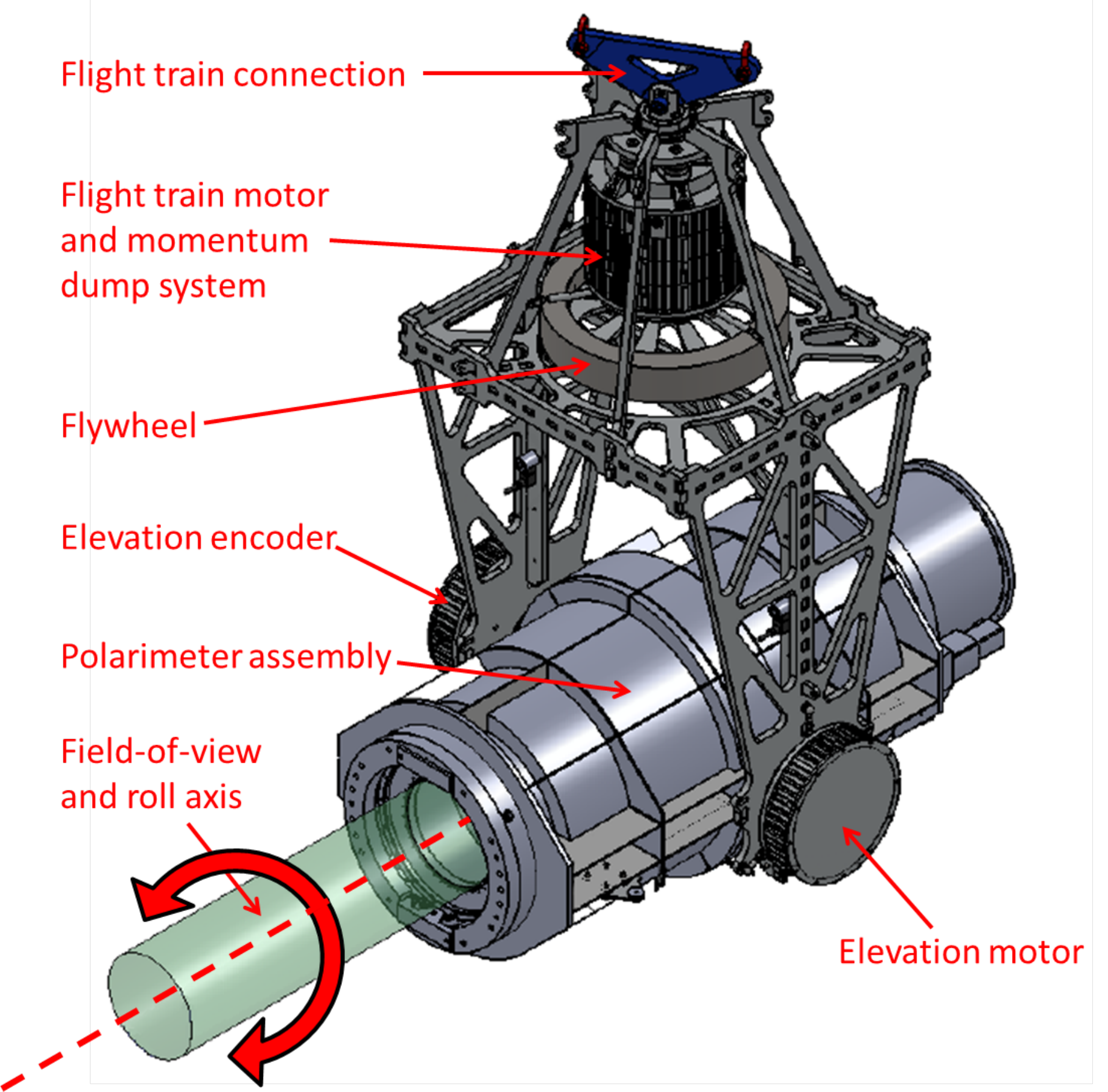}
\caption{\label{Gimbal sketch}Sketch of the gimbal unit assembly housing the polarimeter. The framework stands \mbox{$\sim$2 m} tall and has a mass of \mbox{$\sim$300 kg}, excluding the polarimeter.}
\end{figure}

Celestial targets are acquired and tracked using a custom attitude control system and gimbal unit assembly, designed and developed by DST Control, Sweden. In order to prevent degradation of the polarimetric response due to shadowing effects from the collimators, the pointing is required to be better than 5\% of the instrument field of view, i.e., within \mbox{$\sim$0.1$^{\circ}$}. Coarse azimuthal pointing is achieved using a flight train motor coupled directly to the interface to the balloon rigging. A flywheel provides fine-tuning in azimuth, and excess momentum can be transferred to the balloon through a momentum dump system that controls the feedback to the flight train. Brushless direct drive torque motors are used throughout, with a configuration providing excellent heat dissipation and electromagnetic compatibility. A detailed description of the attitude control system hardware and design philosophy is presented in~\cite{JES reference}. The gimbal unit assembly is shown in Figure~\ref{Gimbal sketch}.

Pointing solutions and motor control feedbacks are calculated based on several attitude inputs. An encoder mounted on the polarimeter axle provides the telescope elevation.~For the azimuth, the~backbone of the system is a differential GPS which, utilizing a \mbox{10 m} baseline, can provide the desired pointing accuracy in itself. A three-axis magnetometer serves as a backup in case of a GPS failure. In the upgrade from PoGOLite to PoGO+, a sun tracker was additionally included. Using a two-dimensional position-sensitive device inside a free-moving gimbal, which automatically locates and follows the sun, this tracker can provide the instrument's azimuth %check
based on its location and the position of the sun. These sensors provide absolute information for instrument pointing. Relative feedback is provided by an optical star tracker camera, mounted co-axially with the polarimeter. For~each X-ray target, an optical guide star is defined.~The camera identifies the pixel coordinates of the guide star and matches these with the expected coordinates, calculated in real time based on the instrument position, altitude, and time of day, correcting for the field rotation of the star field as time goes by. In this tracking mode, motor feedbacks are generated based on the difference between the expected and the observed pixel coordinates of the relevant star.~An inclinometer mounted on the gimbal unit assembly provides information on the gondola pitch and roll.~In addition, a~micro-mechanical inertial measurement unit is used, providing an additional layer of redundancy.~The unit is continuously augmented by inputs from the other available sensors and can, in case other attitude sensors fail, retain attitude information for some time, until the remaining systems become operational again.

\begin{figure}[H]
\centering
\includegraphics[width=.85\textwidth]{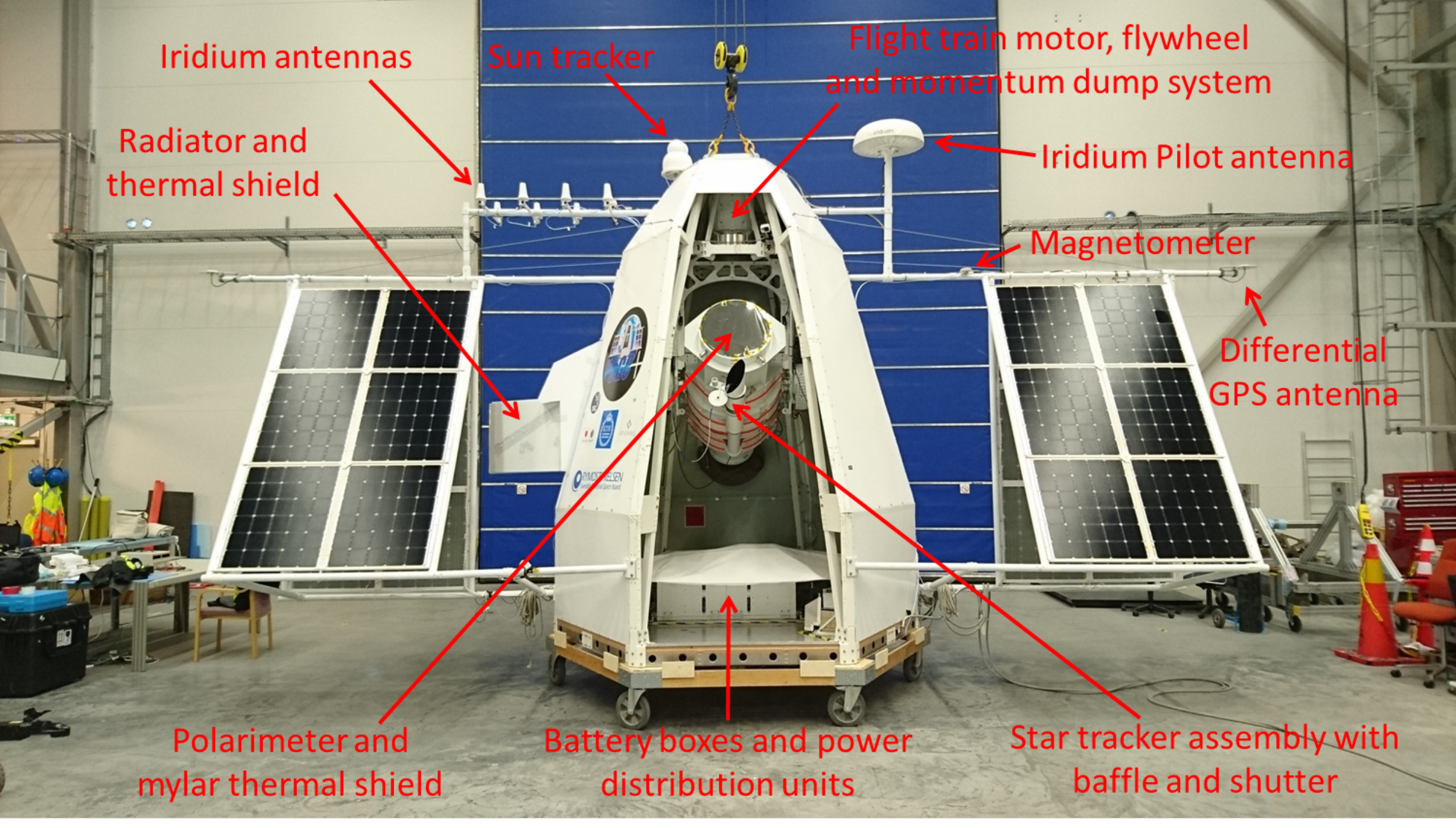}
\caption{\label{PoGO+ payload}The PoGO+ payload and its various constituents.~Crash pads, ballast hoppers, and line-of-sight antennas mounted underneath the gondola are not shown. One panel of the gondola, the front where the polarimeter is looking out, has been removed, showing components inside.  The structure is \mbox{$\sim$4 m} tall and the widest part, at the GPS booms, is \mbox{$\sim$10 m}. For flight, the total suspended weight is almost two tonnes, which includes \mbox{450 kg} of ballast.}
\end{figure}

The polarimeter and gimbal unit assembly are housed inside a gondola structure designed and manufactured by SSC Esrange.~The structure provides thermal shielding, mechanical stiffness for pointing stability, and protection during launch and landing. It consists of a six-sided light-weight framework with aluminum ribs and honeycomb composite plates. Solar panels and booms for the GPS and communication antennas are also affixed onto the structure, as is a radiator for expelling heat from the polarimeter electronics, using a fluid-based cooling system.~During assembly and disassembly, the gondola can be separated into an upper part, housing the polarimeter, and a lower part, which contains batteries, communications equipment, and ancillary hardware. The assembled PoGO+ payload is shown in Figure~\ref{PoGO+ payload}.

%%%%%%%%%%%%%%%%%%%%%%%%%%%%%%%%%%%%%%%%%%

\section{PoGO Flight Campaigns}
\label{PoGO flight campaigns}

PoGO campaigns are conducted from the Swedish Space Corporation (SSC) Esrange Space Center in Northern Sweden (\mbox{67.89$^{\circ}$ N}, \mbox{21.08$^{\circ}$ E}). During the summer, circumpolar winds at a \mbox{$\sim$40 km} altitude move in the westerly direction, allowing flights from Esrange to Canada. The launch window opens around 1 July, when the angular separation between the Crab and the sun exceeds \mbox{15$^{\circ}$}, and closes during August, when the circumpolar vortex starts disintegrating. There have been four campaigns to date, resulting in three launches.

The year 2011 saw the first test launch of the PoGOLite ``Pathfinder.'' A leak of helium from the balloon forced the flight to be terminated prematurely, resulting in a landing near Nikkaluokta, approximately \mbox{100 km} to the west of the Esrange Space Center. Float altitude was never achieved, so scientific measurements could not be conducted. At landing, only minor damage was incurred, and~the polarimeter array could easily be repaired from spare detector units.

A second launch attempt was foreseen for the summer of 2012.~Due to unfavorable weather conditions, the payload remained grounded for the duration of the launch window.

A successful launch of the PoGOLite ``Pathfinder'' took place on 12 July 2013. For this particular launch, an overflight agreement had also been secured from Russian authorities, opening up the possibility for a full polar circumnavigation, concluding with a termination somewhere over Scandinavia. In the end, due to a predicted drift in the trajectory towards higher latitudes, whereby landing in Scandinavia would no longer be certain, the flight was terminated over Russia on 25 July, resulting in a landing near Norilsk, \mbox{$\sim$3000 km} east of Moscow. Details of this flight are presented in~\cite{Payload paper}. Due to issues related to overheating, scientific operations could only be concluded during the first three days of the flight. A statistics-limited measurement of the Crab polarization could, however, be reported~\cite{PoGOLite flight paper}.

\begin{figure}[H]
\centering
\includegraphics[width=.8\textwidth]{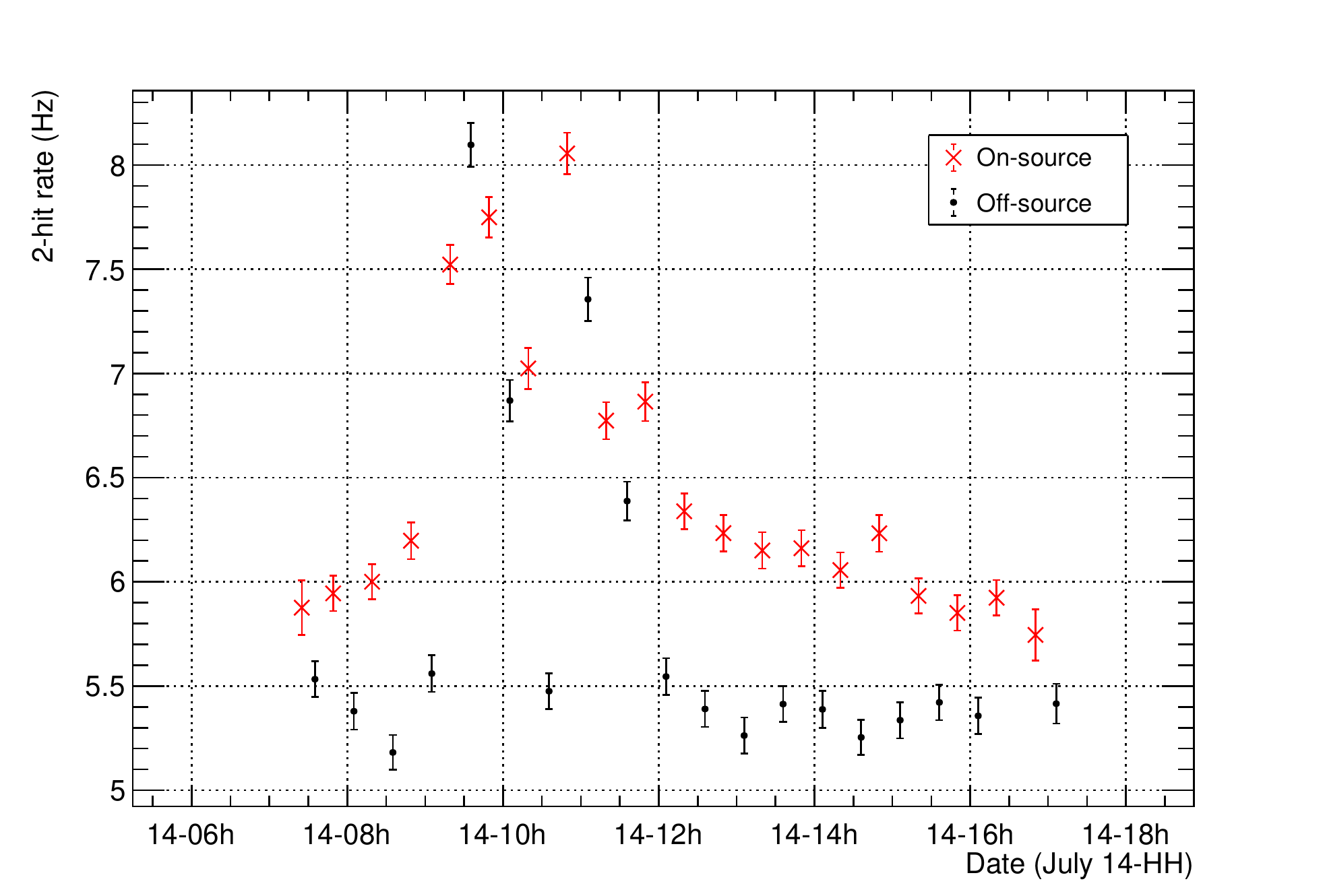}
\caption{\label{Count rate example}Count rate of two-hit events during the Crab observation on 14 July 2016. Apart from the smooth change as a function of time resulting from the sidereal motion of the source across the sky, large fluctuations can be seen both in the on-source (red) and off-source (black) data points. In the absence of interspersed background measurements, this would be indistinguishable from a change in the source rate only, resulting in an incorrectly determined signal-to-background ratio.}
\end{figure}

\textls[-10]{The next launch took place on 12 July 2016, now with the upgraded payload, called ``PoGO+''.} PoGO+ benefited tremendously from data collected during the 2013 flight of PoGOLite. The~polarimeter was extensively redesigned and optimized~\cite{PoGOLite to PoGO+}, including (i) new detector units and improved reflective wrapping yielding higher sensitivity; (ii) improved front-end electronics with a wider dynamic range and a higher waveform sampling rate providing improved efficiency for the pulse shape discrimination and event selection; (iii) additional polyethylene neutron shielding for suppressing the flight background; and (iv) higher sensitivity for the side anticoincidence shield, yielding a lower energy threshold for the background rejection. On the gondola side, an Iridium Pilot antenna had been added, allowing for better monitoring, operations, and data download rates during over-the-horizon communications.~For the attitude information, the aforementioned sun tracker had been added, which simplified observations of the Crab system, where close angular proximity to the sun had previously complicated automatic pointing operations based on feedback from the optical star tracker camera. Of particular importance for the scientific observations was the decision to intersperse background measurements with the on-source observations of the Crab and \mbox{Cygnus X-1}. Throughout the flight, background fields separated by \mbox{5$^{\circ}$} to the east and west of the target were observed, with~transitions to/from the X-ray source occurring automatically approximately every \mbox{15 min}, \textls[-5]{and~with almost equal time spent on source and on background fields.~This strategy allowed the temporal evolution of the background rate and its residual polarization to be studied, and was a prerequisite for correctly determining the in-flight signal-to-background rate, needed for the background-subtraction of the flight data.~The omission of such interspersed background would have resulted in a significant degradation of the systematic precision of the polarimetric results. Figure~\ref{Count rate example} shows an example of signal (on-source) and background (off-source) count rates during a Crab observation, underlining the importance of interspersing the two kinds of observations in this fashion.}

Polarimetric results are derived from the azimuthal scattering angle distributions of these two-hit events. Recorded scattering angles are corrected for the rotation of the detector array throughout the measurement, as well as for the instantaneous gondola roll, sampled at \mbox{4 Hz}. While distributions for both on-source and off-source measurements are first inspected in terms of \mbox{$\chi^2$} per degrees of freedom goodness-of-fit with respect to \mbox{180$^{\circ}$} (polarization) and \mbox{360$^{\circ}$} (anisotropic background) components, % check
an unbinned Stokes-based analysis is used to determine final results. Background is subtracted using the signal-to-background ratio, following corrections for the slight change in acquisition live-time when observing on-source versus off-source. Flight results are summarized in the next section.

%%%%%%%%%%%%%%%%%%%%%%%%%%%%%%%%%%%%%%%%%%

\section{Scientific Results from PoGO}
\label{Science}
\begin{figure}[H]
\centering
\includegraphics[width=.45\textwidth]{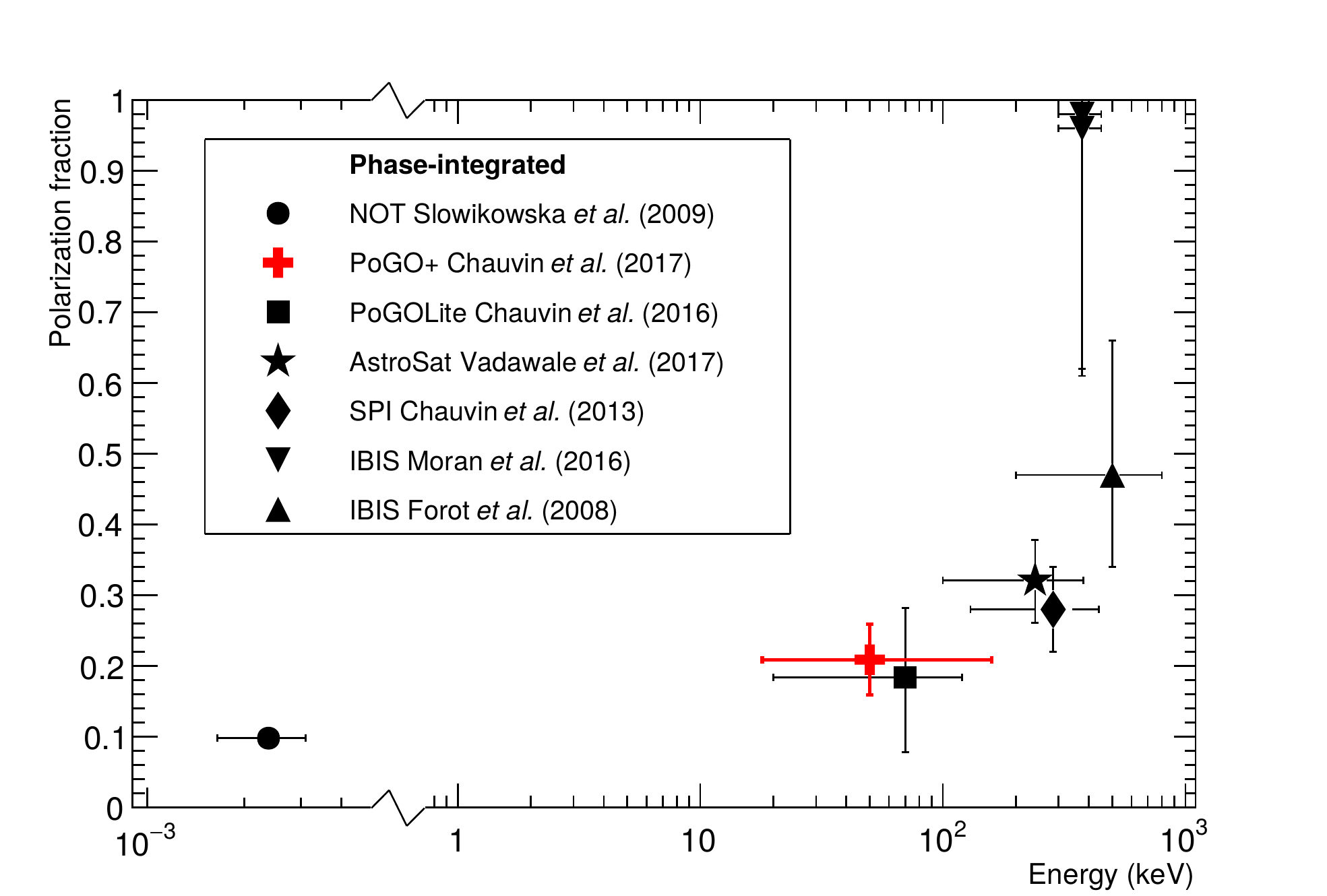}
\includegraphics[width=.45\textwidth]{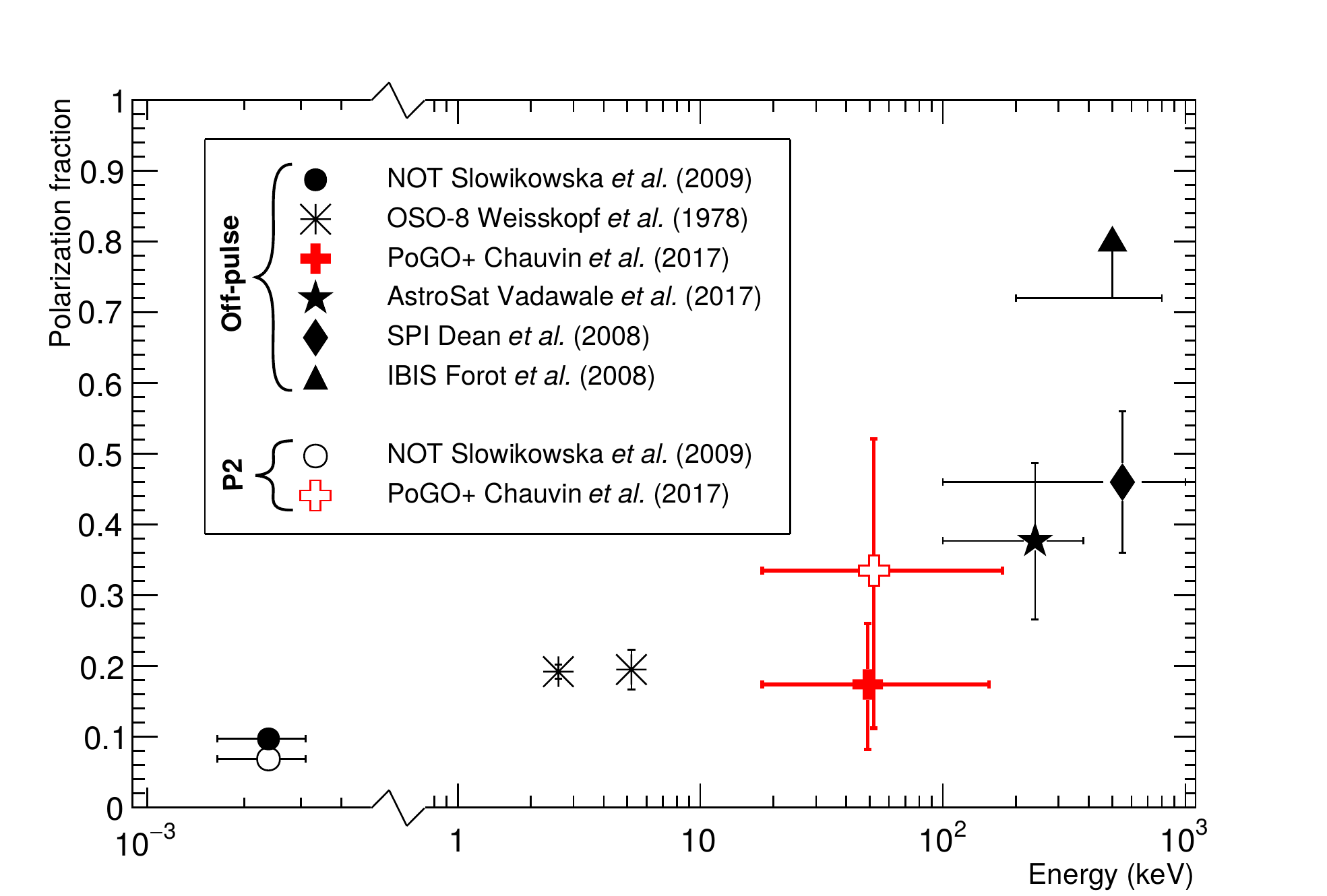}
\includegraphics[width=.45\textwidth]{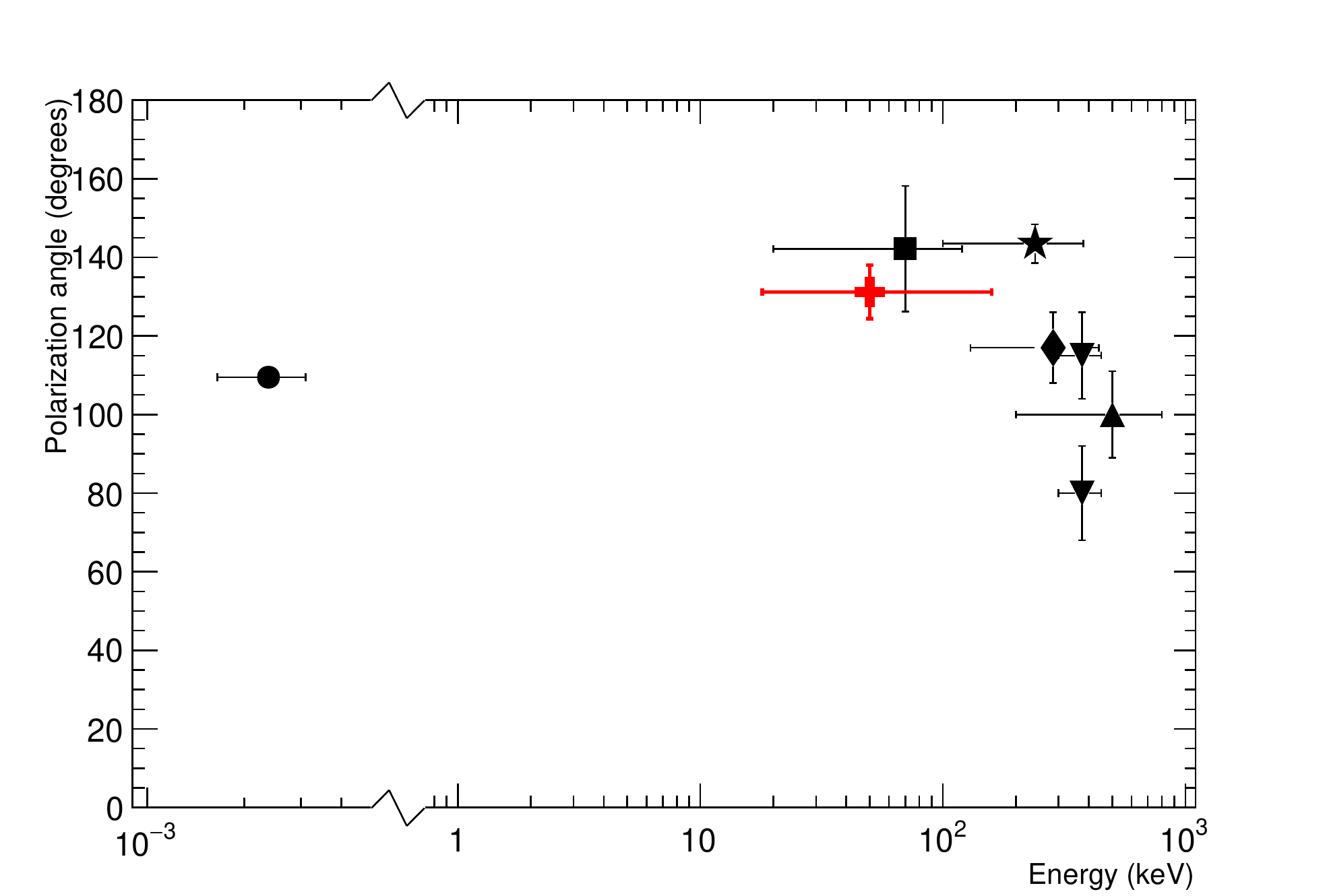}
\includegraphics[width=.45\textwidth]{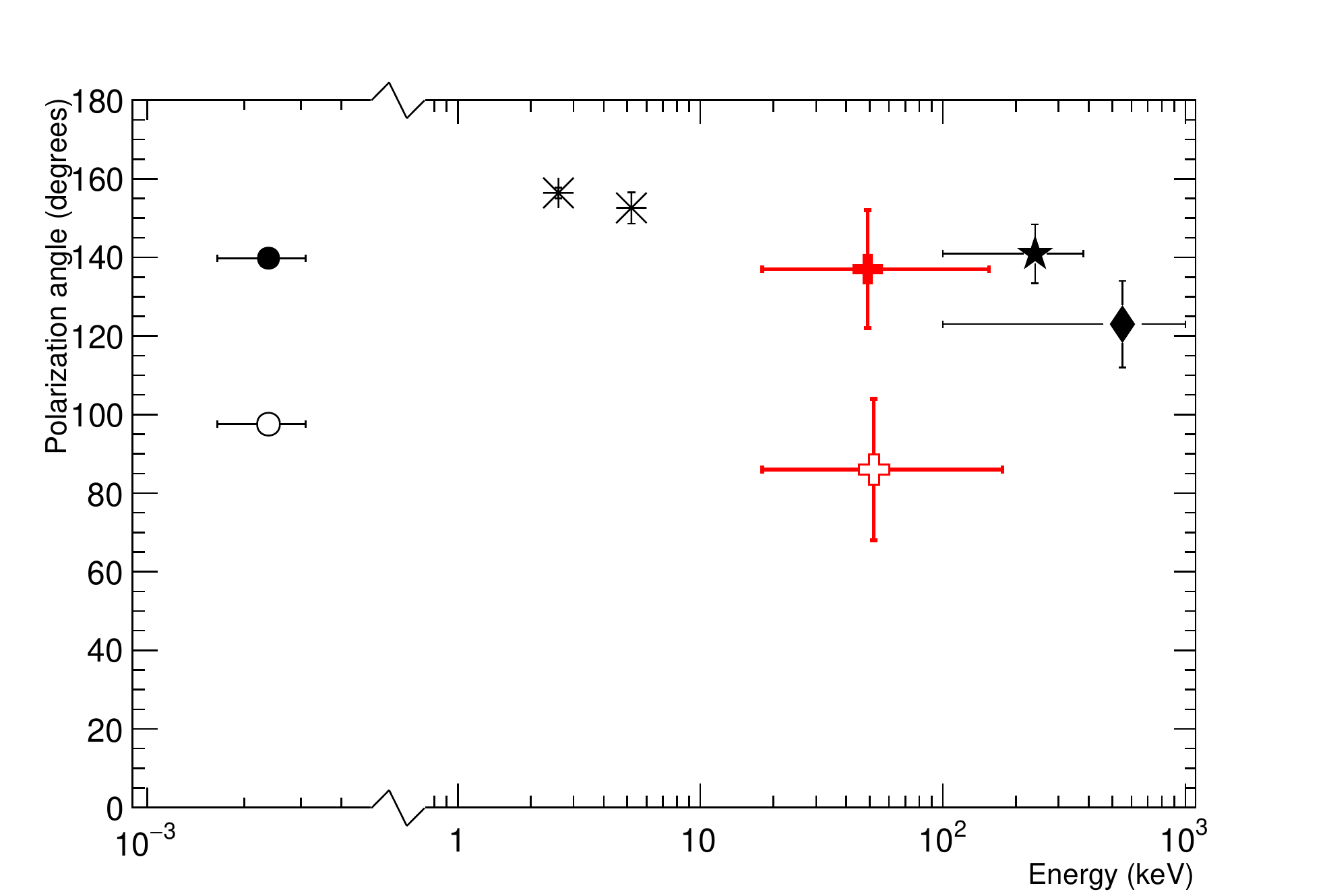}
\caption{\label{Results comparison}Results from PoGOLite and PoGO+ with other polarimetric studies in different high-energy emission regimes, for the phase-integrated Crab measurement (left), and the off-pulse (nebula-dominated) and P2 (off-pulse-subtracted) phase intervals (right), for the polarization fraction (upper panels) and polarization angle (lower panels). Low-energy optical results are also shown.}
\end{figure}

Polarimetric results reported for the high-energy emission of the Crab are limited to a few data points only. At low energies, instruments on the OSO-8 satellite measured polarization at \mbox{2.6} and \mbox{5.2 keV} already in the 1970s~\cite{Weisskopf OSO-8}. For the higher energies of several hundred keV, results are reported from the INTEGRAL SPI~\cite{Chauvin INTEGRAL SPI, Dean INTEGRAL SPI}, and INTEGRAL IBIS~\cite{Moran INTEGRAL IBIS, Forot INTEGRAL IBIS} instruments. The reliability of the INTEGRAL results may, however, be limited by the fact that these instruments were not designed as polarimeters and were not calibrated for polarimetry in ground testing campaigns.~PoGOLite, through the ``Pathfinder'' mission, provided the first polarimetric results in the intermediate energy range, \mbox{20--120 keV}~\cite{PoGOLite flight paper}.~The upgraded PoGO+ instrument later provided an independent study in the range \mbox{19--160 keV}, confirming the findings from PoGOLite, significantly improving the statistical precision and producing a result for the off-pulse (nebula-dominated) and second pulse (``P2'', off-pulse-subtracted) phase regions, respectively~\cite{PoGO+ Crab paper}. Since the PoGO+ results were published, results from the AstroSat mission have also been reported~\cite{Vadawale AstroSat}, in an energy range complementary to that of PoGO+, \mbox{100--380 keV}. Results from these high-energy studies were collected and are shown in Figure~\ref{Results comparison}, where low-energy optical values from the NOT~\cite{Slowikowska NOT} are also included.

The PoGO+ results~\cite{PoGO+ Crab paper} reveal a polarization fraction of \mbox{(20.9 $\pm$ 5.0)\%} for the phase-integrated Crab measurement. This relatively high value indicates synchrotron radiation from a magnetically ordered and therefore compact emission region. The observed polarization angle is \mbox{(131.3 $\pm$ 6.8)$^{\circ}$}, compatible with the pulsar spin axis which is identified as having angle \mbox{(124.0 $\pm$ 0.1)$^{\circ}$}~\cite{Ng pulsar spin axis} projected on the sky. NuSTAR imaging~\cite{Madsen NuSTAR} shows that high energy emission is predominantly coming from more central regions of the Crab system. The alignment of the measured polarization angle with the pulsar spin axis is thus consistent with the toroidal ring region~\cite{Nakamura Crab magnetic field} of the Crab system dominating the hard X-ray emission observed by PoGO+. Spatially resolved measurements from HST~\cite{Moran HST} reveal high levels of polarization from the synchrotron knot and from the wisp structures, exhibiting polarization angles of \mbox{(124.7 $\pm$ 1.0)$^{\circ}$} and \mbox{124$^{\circ}$--130$^{\circ}$}, respectively. A vector map of the full nebula instead shows a peak distribution around polarization angle \mbox{165$^{\circ}$}. Such fine details are not resolved in hard X-rays, but~the agreement with the polarization angle determined by PoGO+ suggests that these features have counterparts also in the energy range of the latter instrument, \mbox{$\sim$19--160 keV}. 

Taken at face value, published polarimetric results for the Crab from different instruments indicate an increase of polarization fraction as a function of energy. However, recent results from AstroSat~\cite{Vadawale AstroSat}  are in agreement with PoGO results~\cite{PoGO+ Crab paper}, which do not show a drastic increase as seen by INTEGRAL IBIS~\cite{Moran INTEGRAL IBIS, Forot INTEGRAL IBIS}. This strain illustrates the difficulty of deriving polarimetric results, polarization fraction in particular, from instruments not calibrated as polarimeters.

%%%%%%%%%%%%%%%%%%%%%%%%%%%%%%%%%%%%%%%%%%

\section{Summary and Outlook}
\label{Outlook}
Unlike satellite-based measurements, scientific ballooning missions allow modifications and upgrades to the payload design to be implemented based on flight data and experience, increasing the chance of success in a re-flight.~In this fashion, PoGO+ has successfully conducted polarimetric observations of the Crab system and \mbox{Cygnus X-1} in the energy range \mbox{$\sim$20--180 keV}, based on experience from the flight of the PoGOLite ``Pathfinder.'' Results presented for the Crab~\cite{PoGO+ Crab paper} are the first in this energy range. Independent results from AstroSat~\cite{Vadawale AstroSat} have been found to be in agreement, strengthening the confidence in the results and countering % check 
a rapid increase of polarization fraction with energy as suggested by some measurements~\cite{Moran INTEGRAL IBIS, Forot INTEGRAL IBIS}. A follow-up to the AstroSat observations, and the claim~\cite{Vadawale AstroSat} of varying polarization properties in the Crab off-pulse region, has also been conducted~\cite{PoGO+ AstroSat rebuttal}, wherein PoGO+ data has been revisited, allowing this claim to be addressed. For~\mbox{Cygnus X-1}, PoGO+ results~\cite{PoGO+ Cygnus paper} allow the emission geometry predicted by different models to be constrained.

What is currently being investigated is the possibility of an energy-dependent analysis of the polarization data, for the Crab and/or for \mbox{Cygnus X-1}. As PoGO+ is optimized for polarimetry and not calorimetry, this would be a Monte-Carlo-driven study, relying on the previously benchmarked simulation results~\cite{PoGO+ calibration paper}.

There are currently no plans to re-fly PoGO+, but the experience from the PoGO mission is used for the development of a new instrument, called SPHiNX (``Satellite Polarimeter for High eNergy X-rays'')~\cite{Fei SPHiNX}, which has a wide field of view and is intended for gamma-ray bursts.

\vspace{6pt}
%%%%%%%%%%%%%%%%%%%%%%%%%%%%%%%%%%%%%%%%%%
\acknowledgments{This research was supported in Sweden by The Swedish National Space Board, The Knut and Alice Wallenberg Foundation, and The Swedish Research Council. In Japan, support was provided by the Japan Society for Promotion of Science and ISAS/JAXA. SSC are thanked for providing expert mission support and launch services at Esrange Space Center. DST Control developed the PoGO+ attitude control system under the leadership of J.-E. Str\"omberg. Contributions from past collaboration members and students are acknowledged.}
%\acknowledgments{All sources of funding of the study should be disclosed. Please clearly indicate grants that you have received in support of your research work. Clearly state if you received funds for covering the costs to publish in open access.}

%%%%%%%%%%%%%%%%%%%%%%%%%%%%%%%%%%%%%%%%%%
\authorcontributions{
%\hl{main text}}   %please provide complete information
  M.F., V.M., M.P. and H.T. attended the Workshop on behalf of the Collaboration. The manuscript was prepared by M.K. All authors have read the manuscript and contributed comments.}
%MK: Information has been provided.

%%%%%%%%%%%%%%%%%%%%%%%%%%%%%%%%%%%%%%%%%%
\conflictsofinterest{
%\hl{main text}}
The authors declare no conflict of interest.}
\end{document}